\documentclass{article}

\usepackage{amsmath,amssymb}

\usepackage{graphicx}

\usepackage{url}

\title{Beyond Spikes: Neural Codes and the \\ Chemical Vocabulary of Cognition}

\author{Romann M.\ Weber}

\date{May 10, 2010}

\begin{document}

\maketitle

\begin{abstract}
In this paper, I examine what I refer to as the \emph{spike doctrine}, which is the generally held belief in neuroscience that information in the brain is encoded by sequences of neural action potentials.  I present the argument that specific neurochemicals, and \emph{not} spikes, are the elementary units of information in the brain.  I outline several predictions that arise from this interpretation, relate them to results in the current research literature, and show how they address some open questions.
\end{abstract}


\section{Introduction: The Spike Doctrine}

One of the fundamental achievements that established neuroscience as a viable and distinct discipline was the development of the \emph{neuron doctrine}, which states that neurons are anatomically distinct cells that serve as the basic computational units in the brain.  In truth, this was not the original statement of the doctrine, which was originally more concerned with the discrete versus continuous character of the brain's neural network, but this is how it is generally regarded in its present form \cite{Bullock}.  Still, it is a concept that continues to evolve \cite{Gold}.

Folded into the current interpretation of the neuron doctrine is what I will call the \emph{spike doctrine}, which is something so firmly ingrained in current neuroscientific theory that one often finds some version of it on or near the first page of any text on the subject.\footnote{The reader is referred to, for instance, \cite{Coolen, DA, GK2002, Harvey, Koch, Rieke, Scott, Trappenberg} among many others.}  A rather poetic statement of this doctrine is given by Rieke, \textit{et al.} \cite{Rieke}, who write:
\begin{quote}
[Our] perception of the world is constructed out of the raw data sent to our brains by our sensory nerves, and in each case these data come in the same standard form---as sequences of identical voltage pulses called action potentials or ``spikes.'' \ldots \space Spike sequences are the language for which the brain is listening, the language the brain uses for its internal musings, and the language it speaks as it talks to the outside world.
\end{quote}

A somewhat more prosaic treatment comes from Gerstner and Kistler \cite{GK2002}, who write:
\begin{quote}
We think of a neuron primarily as a dynamic element that emits output pulses whenever the excitation exceeds some threshold.  The resulting sequence of pulses or ``spikes'' contains all the information that is transmitted from one neuron to the next.
\end{quote}

A second part of the doctrine concerns the assumptions that are made regarding \emph{how} these spikes encode information about the world.  Again, from Gerstner and Kistler \cite{GK2002}: 
\begin{quote}
Since all spikes of a given neuron look alike, the form of the action potential does not carry any information.  Rather, it is the number and the timing of spikes which matter.
\end{quote}
They go on to summarize the doctrine quite explicitly: ``The action potential is the elementary unit of signal transmission.''

Almost without exception, every theoretical effort to consider cognition and information processing in the brain is built upon this doctrine.  It is the aim of this paper to examine this long-held belief in neuroscience and, ultimately, to begin building a case for overturning it.

\section{Why Spikes?}

\subsection{The Electric Brain}

With the proper equipment, it would not take one long to determine that something peculiar is going on inside the brain.  Indeed, the brain is the exemplar structure of the body electric.  Although one does find steady electrical signals issuing from the heart, the brain is unique in its level of galvanic chatter.  It is perhaps no wonder that the prevailing belief is that crackling inside this tangle of electrical activity is the very language of thought.

In addition to seeming very special and \emph{brain-like}, this electrical activity also has the advantage of being relatively easy to measure, at least at a coarse level.  Monitoring the activity of an \emph{individual} cell is much more difficult, however.  It currently requires getting into the skull and placing an electrode in or near the cell of interest, a procedure typically reserved for certain surgical patients and hapless laboratory animals.

That neurons produce any measurable electrical activity at all seems a minor miracle in itself.  Neurons do not use electricity the same way a household appliance does.  Rather, neurons regulate the flow and concentration of charged ions on either side of their cell membranes.  The difference in charge between the inside of the cell and the outside is known as a \emph{membrane potential}, and a spike is a rapid shift in this potential that propagates down the length of the nerve.  The arrival of a spike at the end of the neuron triggers the release of stored chemical \emph{neurotransmitters}.

The most immediate effect these chemicals have is to open gated ion channels on the postsynaptic neuron, thereby changing its membrane potential.  Causing the influx of positive ions increases the cell's chance of firing a spike (in which case the responsible neurotransmitter is called \emph{excitatory}), and causing the influx of negative ions decreases it (in which case the transmitter is called \emph{inhibitory}).  In short, by releasing neurotransmitters, one cell can influence whether the cells it is signaling fire spikes of their own.

This easy-to-state relationship, highly influenced by the early work of McCulloch and Pitts \cite{MP}, is at the core of artificial neural network (ANN) research, much of the theory for which can be summarized in the single equation 
\begin{equation}
	y_i = f \left( \sum_{j=1}^n w_{ij}x_j - \theta_i \right). \label{ANN}
\end{equation}
Here $y_i,$ the ``state'' of neuron $\nu_i,$ is determined by applying some (often nonlinear) function $f$ to the collected states, $x_j,$ of the $n$ neurons connected to it multiplied by the ``strength'' of those connections, represented by the \emph{weight} term $w_{ij}.$  The $\theta_i$ term describes the \emph{threshold} for the neuron $\nu_i$ \cite{HK2001}.\footnote{Not all ANN models use the exact form of equation \eqref{ANN} to compute neural states, but the vast majority---and virtually all simple models---use it or a close relative.  It should also be noted that the output, $y,$ would be an input, $x,$ to some other neuron unless it was the member of the network's \emph{output layer}.}  An example of a general network---although not an ANN---is shown in Figure 1.

\begin{figure}[t]
\begin{center}
\includegraphics[scale=0.55]{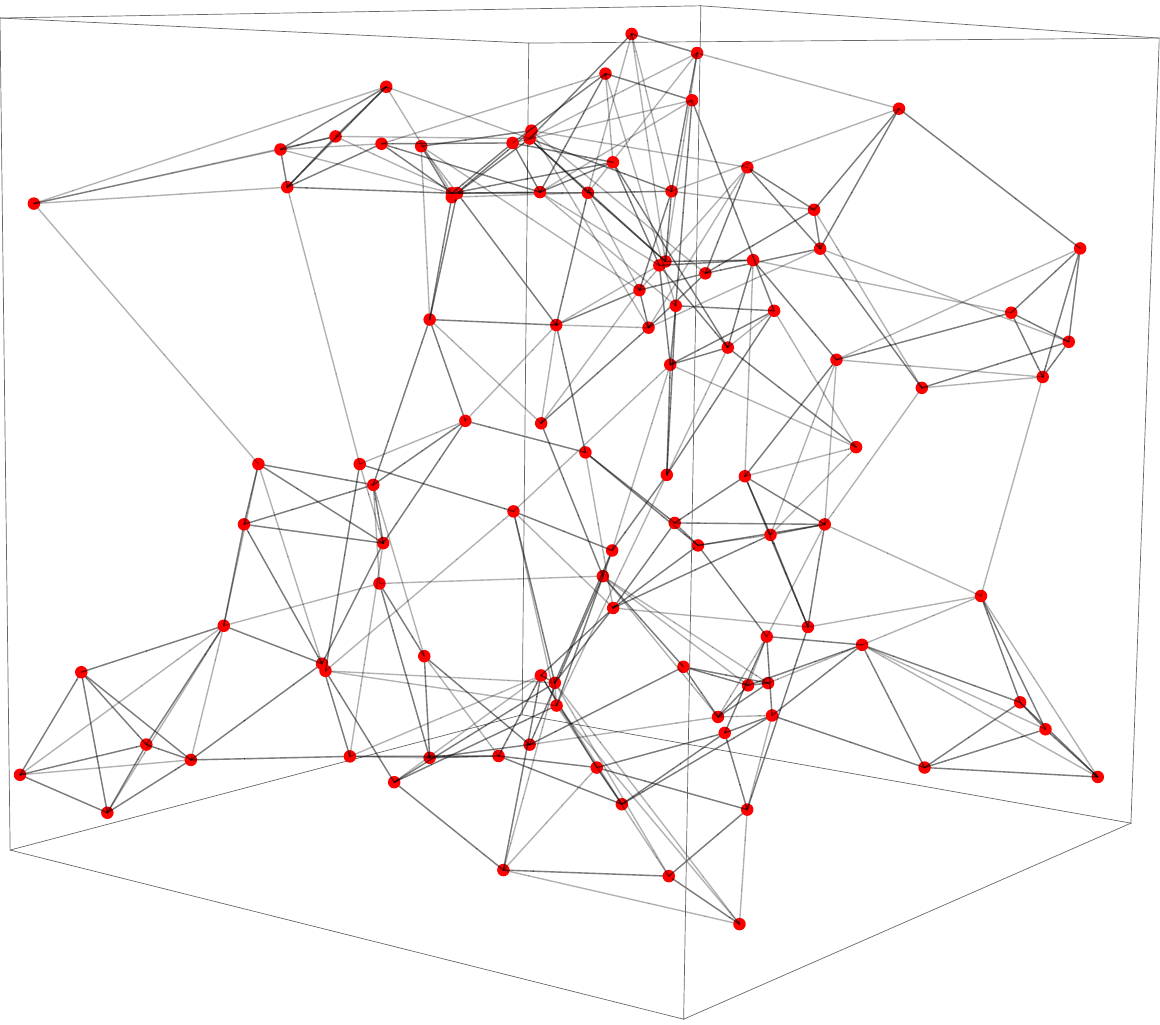}
\caption{An example of a 3D network.  Each red point is a \emph{node}, which would serve as a ``neuron'' in an ANN.  Information is exchanged via the connections (\emph{edges}) between the nodes, governed by a connection strength described by the \emph{weight}, $w.$  Image generated from code found in \cite{HW}.}
\end{center}
\end{figure}

In some models, the state of the neuron may simply be 1 or 0, which could be considered to correspond to whether it is firing or not.  The state can also be a firing \emph{rate} or even a membrane potential.  A considerable portion of ANN research, which we will refer to by the more general term \emph{connectionism}, does not concern itself too much with biological realism, so the ``neuron'' states do not have to correspond to anything an actual cell has to deal with.  Even in these cases, though, some of the biological language is preserved.  So, if a neuron is considered to be inhibitory, its connection weight to postsynaptic cells will generally be negative; it will be positive in the excitatory case.

Artificial neural networks are interesting objects, and they can be ``trained'' to perform a variety of important tasks, from pattern recognition and classification to function approximation.  Further, it has been shown that certain finite neural networks---in fact, neural networks with less than one-millionth the number of neurons in the human brain---can simulate a universal Turing machine \cite{SS91}.  In essence, this means that such a network---given enough time---can compute anything that \emph{can} be computed.\footnote{A Turing machine is not an \emph{actual} machine but rather a theoretical construct that serves to illustrate what computers can and cannot do.}  Other research has suggested enormous computational potential for spiking neural network models.\footnote{See, for instance, \cite{Maass}.}  

Artificial neural networks ``learn'' to perform certain functions through a training regimen whose purpose is to alter the values of the synaptic weights, $w,$ until the inputs to the network produce outputs that are as close as possible to \emph{target} outputs.\footnote{These input-output pairs used for learning are often referred to as the \emph{training set}, and the method by which the weights are adjusted is referred to as a \emph{learning rule}.}  This is one example---of many---where an ANN deviates from biological realism, since a \emph{real} neural network cannot be expected to have access to whatever the ``target'' outputs are supposed to be.

Even so, the basic principle of learning for an ANN seems to apply in the brain.  \emph{Synaptic plasticity}, a molecularly and genetically mediated process that adjusts the strength of connections between neurons, is thought to be the chief component behind learning and memory \cite{Kandel}.  

\subsection{The Neural Code}

Current theory states that spikes pulsing through the nerve network in the brain represent and convey information, and the adjustment of the connections between nerves gives rise to memory, learning, problem solving, and everything else that makes us intelligent.  The manner in which this information is represented by spike sequences is often referred to as the \emph{neural code}.  There is considerable debate as to the exact form of this code, especially with regard to whether the precise \emph{timing} of individual spikes or the average \emph{rate} of spikes carries the necessary information \cite{Butts, DA, Rieke}.

Temporal coding, rate coding, and population coding (which takes into account the collective activity of multiple neurons\footnote{Population coding is a term that can technically include \emph{correlation coding}, in which information is encoded by comparing the timing of one neuron's spike sequence to another's.}) are the prevailing theories for how information is encoded in the brain.  Indeed, if information is conveyed by \emph{spikes}, then the neural code \emph{must} be described by at least one of these methods.  But, as there is at least some evidence to support each of these methods, deciding among them is difficult.

It is worth pointing out that rate and population coding have enjoyed considerable success, particularly in vision research and the study of perception and action.  For instance, the ``preferences'' of individual neurons for highly specific visual features seem to be indicated by their firing rate, and calculations using the concerted activity of a small population of neurons in a monkey's motor cortex can predict its arm movements with high accuracy \cite{GIM}.

Researchers who believe that spike sequences encode information in the brain have a fair amount of encouraging evidence on their side.  As we have covered earlier, neural spiking is mediated by various neuroactive chemicals.  Clearly, then, if the information is encoded in action potentials, then it must be the case that these chemicals function in the \emph{service} of the spikes.  We will soon examine the possibility that it is the other way around.

\section{Why \emph{Not} Spikes?}
\subsection{The Brain as Statistician}

Most experiments that investigate the neural code do so by monitoring the response of a certain cell to a controlled stimulus.  Let us say that over some interval of time, we make a note of the times at which the cell fires a spike.  We can denote this sequence by $\{ t_i \}.$  This gives us information not only about the precise timing of each spike but also about the average firing rate, which we can obtain by counting all the elements in the sequence and dividing by the amount of time we were monitoring the cell.

Scientists looking for the neural code would like to identify $P[ \{ t_i \} | s(t) ],$ which describes the \emph{conditional} probability of forming the spike sequence $\{ t_i \}$ given that the stimulus $s(t)$ has occurred.\footnote{Phrasing the neural code in terms of probabilities may seem like equivocation.  However, those looking for a more definite, prescriptive approach should bear in mind that a probabilistic statement can also encompass ``certain'' events of probability 1.}  The neuron, however, has essentially the opposite problem, as it would like to evaluate $P[ s(t) | \{ t_i \} ],$ which describes the probability of a certain \emph{stimulus} given the spike train \cite{Rieke}.

These probabilities are related by a classic result known as Bayes's theorem, which takes the form 
\begin{equation}
	P[ s(t) | \{ t_i \} ] = \frac{P[ \{ t_i \} | s(t) ] P[s(t)]}{P[ \{ t_i \}]}. \label{Bayes}
\end{equation}
This result states that the probability of a particular stimulus \emph{given} a specific observation (the spike train, $\{ t_i \}$) is proportional to the probability of the observation given the stimulus---which makes intuitive sense---but it is controlled by the ratio of the probability of the stimulus, $P[s(t)],$ to the probability of the observed spike train, $P[ \{ t_i \}].$

Now, while a neuron may conceivably know what its own response history is, which would allow it to evaluate the response probability, where does its knowledge of the \emph{stimulus} probability come from?  The implication is that even if one has complete knowledge of the scheme for \emph{encoding}, anything short of an \emph{exact} prescription for this encoding results in uncertainty when \emph{decoding}, that is, when interpreting an observation.  Much current theory seeks to address this uncertainty.

\subsection{An Explosion of Uncertainty}

The vast majority of neural network models assume that when a neuron fires, the message that it has fired is sent faithfully.  This is not the same as saying that some models do not assume there to be \emph{noise} in the system.  The difference is that the noise in those models is phrased in terms of random nerve activity that does not encode anything meaningful.  But there is another form of ``noise'' in real neural systems that may prove more troublesome.

Before we develop this point, however, let us return to the connectionist paradigm for a moment.  A neuron's spiking pattern is influenced---\emph{caused}, we might say---by the spiking patterns of ``upstream'' neurons that connect to it based on how strong those connections are.  With this in mind, another way of writing equation \eqref{Bayes} is 
\begin{equation}
	P[ s(t) | \{ t_i \} ] = \frac{P[ s(t), \{ t_i \} ]}{\sum_{\mathbf{x}(t)} P[ \{ t_i \}| \mathbf{x}(t)] P[\mathbf{x}(t)]}, \label{prob_2}
\end{equation}
where $\mathbf{x}(t)$ is a time-varying vector that collects all the spike sequences of the upstream neurons.  It looks considerably different from the expression in equation~\eqref{Bayes}, especially since it now invokes the joint probability $P[ s(t), \{ t_i \} ],$ which considers the pairing of all possible stimuli and responses.  It also requires that we consider the response $\{ t_i \}$ in relation to all possible inputs $\mathbf{x}(t).$  But if the rule for the response is as straightforward as equation \eqref{ANN}, this new requirement need not cause \emph{too} much additional difficulty for the cell.  That is, it need not unless $P[\mathbf{x}(t)],$ the output of the upstream neurons, is \emph{itself} uncertain.

This turns out to be the case.  Although there is some evidence to suggest that changes in one neuron's membrane potential can have \emph{direct} influence on a neighboring neuron's potential, this cannot be employed as a general method for neural signaling.  After all, neurons routinely communicate with other neurons far outside the influence of the extraordinarily weak electric and magnetic fields that they generate.  The only way for one neuron to know that another neuron has fired is to receive a chemical signal from it.\footnote{An exception occurs in so-called \emph{electrical synapses}, which communicate via direct ion exchange through \emph{gap junctions}.  The current viewpoint, supported by various models, is that gap junctions help enforce \emph{synchronous} firing among oscillating neurons \cite{KE}.}

Neurons do not reliably release neurotransmitter with each spike.  Not only does the transmitter release probability vary widely \emph{among} neurons, this probability is also variable for \emph{each} neuron and can change over time \cite{Branco}.  One estimate is that, on average, synapses have a transmitter release probability of about 30 percent \cite{Linden}.  If we replace $\mathbf{x}(t),$ the upstream spiking record, with $\mathbf{\xi}(t),$ the firing indications that our neuron actually \emph{receives}, then equation \eqref{prob_2} becomes 
\begin{equation}
	P[ s(t) | \{ t_i \} ] = \frac{P[ s(t), \{ t_i \} ]}{\sum_{\mathbf{x}(t)} \sum_{\mathbf{\xi}(t)} P[ \{ t_i \}| \mathbf{x}(t)] P[\mathbf{x}(t)|\mathbf{\xi}(t)] P[\mathbf{\xi}(t)]}, \label{prob_3}
\end{equation}
which introduces a tremendous amount of additional uncertainty (and effort) into the problem.  Indeed, establishing the new term $P[\mathbf{x}(t)|\mathbf{\xi}(t)],$ which can be interpreted as a complete description of the release probabilities of all cells in a neural circuit, is itself a task almost on par with the difficulty of the original signal-interpretation problem stated in equation \eqref{Bayes}.

The variability in transmitter release would seem to force us to discard \emph{temporal} coding as a viable general method for representing information in the brain.  And, although \emph{rate} coding remains a possibility---since each input neuron's average observed spiking rate would be scaled by its transmitter-release probability---it would seem to require that each neuron have a complete and up-to-date account of the release probabilities for the synapses it shares with upstream neurons.  Our neuron's signal-identification problem has become fraught with ambiguity.

\section{Signals in the Cellular Domain}

\subsection{Neuroactive Molecules}

We have already discussed one role of neurotransmitters in signaling among cells, namely the manner in which they influence the downstream firing of action potentials.  Neurotransmitters lie within a broader class of neuroactive chemicals that include chemicals known as \emph{neuromodulators}, which, as the name implies, modulate a cell's response to other chemical signals.  Neuromodulators include neuropeptides, which are chains of amino acids; certain gasses dissolved in the intercellular medium, such as nitric oxide; and some products of fatty-acid metabolism.

For a chemical to be classified as a neurotransmitter, it must satisfy certain criteria \cite{GIM, vBD2006}:
\begin{enumerate}
	\item It must be manufactured by the neuron and released into the synapse in response to an action potential.
	\item It must induce effects on the postsynaptic cell, mediated by transmitter-specific receptors.
	\item There must be a mechanism for inactivating or removing the substance from the synapse after release.
	\item Experimental application of the substance to nervous tissue must produce effects similar to those induced by the naturally occurring transmitter.
\end{enumerate}

We are familiar with some of the effects described in the second point, as we know that some neurotransmitters are excitatory (the most common of which is glutamate) and some are inhibitory (the most common of which is $\gamma$-amino butyric acid, or GABA).  The receptors for these chemicals control gates that regulate ion flow, and such receptors are collectively called \emph{ionotropic}.

There is a second class of receptors, however, known as \emph{metabotropic} receptors.  Once triggered, these receptors initiate complex chemical cascades that generate additional signals \emph{within} the neuron \cite{vBD2006}.  The ultimate recipient of these intracellular signals is the neural genome, as genes are selectively expressed via transcription factors assembled and activated in response to the cascade \cite{Stahl}.
	
\subsection{The Genomic Response}

Like almost all cells in the body, every neuron has a complete copy of the genome in its nucleus.  And, also like other cells, only a small fraction of the genes within each neuron's copy of the genome are expressed at any time \cite{Linden}.

It is well known that during development, cells differentiate by expressing a specific subset of genes, regulated by a chemical process that is not completely understood.  One pattern of gene expression leads a young cell toward a fate of bone, for instance, while another pattern leads to brain.  As gene expression controls not only certain housekeeping duties of the cell but also the \emph{products} it makes, we would expect neurons that manufacture, say, dopamine to show a different pattern of gene expression than a neuron that manufactures acetylcholine.  This is indeed true, but the differences among neural gene-expression patterns do not end there.

Selective gene expression does not stop after development, after all.  Perhaps the most interesting example in neurons is known as the \emph{immediate-early gene} (IEG) response.  This is a specific pattern of gene expression in the neural genome in response to a stimulus, a reaction so specific as to have been referred to as the ``genomic action potential'' in the literature \cite{Clayton}.

Much recent work has demonstrated this effect in the zebra finch in response to the vocalizations of its fellow finches \cite{Clayton, Dong, Warren}.  Further, this genomic response ceases once the vocalization becomes familiar \cite{Dong}.  A number of theories suggest that this response is necessary for memory consolidation of specific events, but it has also been proposed that the response instead improves the efficiency of memory formation by altering the cell's state to better encode transient, but similar, experiences into long-term memory \cite{Clayton}.

It is known that gene expression and subsequent protein synthesis play a role in memory formation, and inhibiting protein synthesis has been shown to greatly interfere with learning \cite{Igaz}.  One could interpret the role of protein synthesis in memory formation from a strict connectionist viewpoint and insist that it be in the service of the \emph{long-term potentiation} (LTP), or strengthening, of excitatory synapses---namely, adjusting the ``weights'' among interacting cells.  This strict interpretation is challenged, however, by the finding that inhibiting LTP at $N$-methyl-\textsc{d}-aspartate (NMDA) receptors for glutamate does not prevent experimental animals from learning certain tasks \cite{GIM}.\footnote{Blocking LTP does interfere with learning in certain contexts, however.}  As is often the case in biology, the reality appears to be far more complex.

Particularly interesting is the way in which the IEG response appears to be independent of spike activity.  For instance, as reviewed in \cite{Clayton}, we find the following phenomena in zebra finches: Songs that have become habituated by repitition cease to induce a gene response, but this cessation of the gene response occurs \emph{despite} continuing spike activity induced by the stimulus.  Songs of heterospecific and conspecific birds induce similar spiking behavior in the caudomedial neostriatum, but heterospecific songs induce the IEG only \emph{half as effectively}.  And singing induces considerable gene activation in a song-control nucleus in the basal ganglia, but those cells' firing rates show \emph{little to no increase} during singing.

Here, then, we have distinct cellular responses to stimuli that are \emph{not} distinguished by their spiking behavior.  It is not at all clear how a strict connectionist viewpoint could account for these phenomena.

\section{The Chemical Vocabulary of Cognition}
\subsection{A Game of 37 Questions}

Imagine that you are given a printout of a neural spike train and are asked to identify the stimulus that provoked it.  What might your answer be?\footnote{The correct answer is: ``I have no idea.''}

If you are a cognitive neuroscientist, you may try to get some additional information out of your interlocutor.  If you learn, for instance, that the cell is in the occipital lobe, then you can be fairly sure that the stimulus was a \emph{visual} signal of some sort.  But ``a visual signal'' is unlikely to be a winning response in this game, so you still need to know more.  If you learn that the cell is from area V1, then you know a bit more, but still not enough to unequivocally identify the specific feature of the stimulus that the cell is responding to.

Cognitive neuroscience research is concerned with identifying the neural correlates of behavior, perception, action, emotion, and cognition in general.  Much of that research involves locating specific parts of the brain that appear to be involved in producing or influencing these phenomena.  This research has shown consistently that certain parts of the brain are specialized for dealing with certain types of information.  It has also been shown that the activity of individual cells can be highly preferential, responding strongly to very specific types of stimuli and hardly at all to others \cite{GIM}.

Could identifying a stimulus from a spike train be made possible by identifying the cell that produced the spikes?  Could we identify a single cell in the human brain in a game of 20 questions?  If we are restricted to \emph{yes}-or-\emph{no} questions, then we cannot.  It would take, on average, $\log_2 10^{11} \approx 37$ such questions to identify an individual cell among the 100 billion in the human brain.

What is the point of considering such a game?  Recall from section 3.2 that experimental research into the neural code often takes the form of monitoring a cell's response to a given stimulus.  From the discussion that followed, we know that during an experiment, we researchers have more information than the cell does, since we know the stimulus the cell is tasked with identifying.  But we also know something else, namely the specific \emph{location} of the cell being monitored.

This may seem to be a trivial point, but we must keep in mind that the additional context we would need for trying to make sense out of a spike train should also be required by a \emph{neuron} trying to make sense of a spike train.  After all, although a neuron's activity can be highly variable, it is limited to a relatively narrow range, almost never exceeding 1000 Hz (spikes per second) but usually topping out at about 400 Hz \cite{Linden}.  As Harvey writes \cite{Harvey}:
\begin{quote}
The origin and targets of nerve fibers establish information and meaning because signals are similar in all nerve cells.  That is, meaning has to do with the particular neural group, while frequency coding conveys information about the stimulus intensity.
\end{quote}

In other words, even if spikes do encode crucial information, neurons should need to know what neurons are \emph{sending} them in order to properly interpret the incoming signals.\footnote{And we would want the method to be completely general, allowing for communication among \emph{any} neurons in the brain.  If each cell were limited to distinguishing among the 10,000 or so other cells the average neuron has in its network, then it would not be possible to form a relationship with a cell \emph{outside} of that network, since any other cell would possess the chemical ``phone number'' of one already in the network, so no new infomation could be gained.}  The question becomes how one neuron might tell other neurons apart.

\subsection{Chemical Calling Cards}

The brain is a three-pound biochemical laboratory.  To date, well over 100 neuroactive chemicals have been identified, and the list continues to grow.\footnote{One measure of the pace at which this list is growing is the fact that a popular handbook of neuroactive chemicals \cite{vBD2006} grew by 100 pages in the four years between its first and second editions.}  Neurochemicals are often found in well-defined circuits in the brain, such that detecting a certain chemical in a signal gives you considerable information as to the part of the brain from which the signal originated.

We could imagine, then, that the presence or absence of a certain chemical in a signal could be interpreted as the answer to a \emph{yes}-or-\emph{no} question.  As we saw above, it would take about 37 such questions to specify a single cell among 100 billion, meaning that the brain would require the selective, combinatorial expression of 37 neuroactive chemicals in neurons if it wished to chemically encode signal provenance.  Clearly, with over 100 neurochemicals and counting, \emph{specificity} regarding neural signals is a problem the brain need not face.

Anatomical location may not be the main variable of interest to the brain.\footnote{Even so, location \emph{is} important to the brain, as suggested by the multiple location-preserving maps found in the visual and sensorimotor systems.}  But it is clear that the brain possesses a \emph{chemical} means for providing additional information---or context---for a signal.  Since the chemical ``messages'' associated with this additional information would undoubtedly outnumber the distinct types of neurochemicals in the brain, we would expect to see individual cells express \emph{multiple} transmitters if specific information is to be encoded chemically.

Indeed, this is the case.  Not only is it true that neurons express multiple transmitters, but they also do so in highly specific patterns \cite{Black}.  Cotransmission of neurochemicals is now known to be ``the rule rather than the exception'' \cite{Trudeau}, but the physiological purpose of such cotransmission is not understood and is currently regarded as an open question \cite{Trudeau, vBD2006}.

For an example of how chemical context is provided in neural signals, consider replacing the cholinergic neurons in the spinal cord with dopaminergic cells.  A moment's thought should convince the reader that the behavioral differences would be obvious, even if the spiking of the impostor cells were experimentally manipulated to exhibit the \emph{same} spiking behavior of the original cells.  Although this example is a thought experiment, \emph{actual} transplantation of neurons from one brain region to another has produced astounding results.

In one study, serotonergic cells from the mesencephalic raph\'{e} nucleus expressed substance P after transplantation to the hippocampus and striatum,
but this effect was not seen upon transplantation to the spinal cord \cite{Schultzberg}.

It is apparent that a specific region of the brain can be indicated by a chemical signature, one that is so influential as to impose itself on foreign cells when they are introduced.  But the brain is not a device running a fixed program.  If specific information is actually encoded by the chemical signature of neural signals, then it should be necessary that a neuron's transmitter phenotype be allowed to change over time.  This is also true, and it has been shown that neurotransmitter phenotypes are highly plastic and can change in response to activity or environmental signals \cite{Black, Changeux, Potter, Trudeau}.

\subsection{An Expanding Computational Landscape}

Chemical signaling is an ancient technique, and it is still employed as the sole means of communication by our distant, nerveless cousins.  Interestingly, ``neurotransmitters'' such as epinephrine, norepinephrine, and serotonin are contained in sponges, in which there is nary a neuron to be found \cite{Black}.  These putatively subordinate chemicals actually evolutionarily \emph{predate} the cells they are assumed to serve!

Chemical signaling in the cell is a complex process often referred to as \emph{signal transduction}.  The interactions in these transduction networks have the appearance of complex ``neural'' networks, so the question of their computational potential is a natural one.  Indeed, recent opinion has advocated the investigation of the contribution of these chemical networks to neural computation \cite{Black, KC, Le Novere, Thagard}.

Specific studies have indicated that neural biochemical networks are capable of decoding complex temporal patterns of input \cite{Bhalla2002}.  The kinetics of chemical reactions has also been shown to enjoy the same Turing universality demonstrated for certain neural networks \cite{Magnasco}.  Further, it is becoming evident that microRNA plays an important role in these complex intracellular chemical-genetic networks \cite{Schratt06, Schratt09, Warren}.  This is particularly remarkable, since microRNA is produced from \emph{noncoding} DNA, which makes up well over 90 percent of the human genome but was once regarded as ``junk'' \cite{Revest}.

Although the computational complexity of the individual neuron is being increasingly appreciated by researchers, it remains unorthodox to suggest a role for biochemical and genetic components in cognitive computation in a manner that challenges the spike doctrine.  That is, the prevailing view still assumes the basic unit of information signaling to be the action potential, in whose service the chemical and genetic components are still assumed to function.\footnote{An example of this assumption comes from a 2002 article by Paul Thagard, in which he writes [italics mine]: ``Neural-network computational models typically treat neuronal processing as an electrical phenomenon in which the firing of one neuron affects the firing of all neurons connected to it by excitatory and inhibitory links. \ldots \space The role of neurotransmitters and other molecules \emph{in determining this electrical activity} is rarely discussed'' \cite{Thagard}.  On the other hand, a recent example of the rare, ``unorthodox'' view I promote is due to Emanuel Diamant, who discusses the inadequacy of spike trains as the sole means of information transfer, favoring a biochemical alphabet instead \cite{Diamant}.}  It is only the expanding repertoire of this servile function that appears to be appreciated.

It has never been claimed in the neuroscience literature that neuroactive chemicals are unimportant or meaningless.  But we are beginning to get some indication of the deeper role they may play in the representation of information in the brain---indeed, a \emph{central} role.  Earlier we considered the function of neuroactive chemicals as being in the service of spikes.  So how might it be the other way around?

If we believe that information is represented in the brain by patterns of neural electrical activity, then we must also consider the implications of that belief.  Namely, we should convince ourselves that the evolutionary advent of the nerve and the action potential coincided with the abandonment of \emph{specific} informational roles for biological molecules, molecules that had served a symbolic, informational purpose in our preneural ancestors.  We should also ask why the action-potential paradigm, which clearly must be rich enough to deliver all the complexity of consciousness, is not used exclusively throughout the body as a means of information transfer.

The fact is that while all cells in the body communicate chemically---and, indeed, also possess the genetic instructions and therefore the potential for electrical communication---only the cells that require fast communication outside of their immediate physical neighborhoods employ the action potential.  This suggests that the action potential is simply a means for \emph{delivering} the message---and an expensive one at that \cite{Lennie}---but not the message itself.\footnote{Considerations of the cost of encoding information allow us to make certain experimental predictions.  Assuming that it is more costly to express multiple transmitters, we might expect to see the the variability of transmitter expression---with a tendency toward increasing cotransmission---change as we progress through a neural circuit.}

Why, then, do studies of spiking seem to provide so much information?  The answer may come from an example drawn from current events.  Many news reports in the years since September 11, 2001, have reported on ``chatter'' among cells of terrorists, with the suggestion being that the dynamics of this chatter indicates something about the threat posed by these cells.  This is, of course, information that is completely independent of the \emph{content} of that chatter.  The messages, in other words, are unknown; we simply know that more or fewer are being sent.  Such may also be the case with the very different---but even more mysterious---cells that we have been discussing in this paper.

\section{Neural Knowledge}

What does a neuron \emph{know}?  As we have discussed earlier, neural responses to stimuli can be very specific.  Some neurons respond preferentially to faces, and some respond preferentially to \emph{specific} faces.  But does this mean that there is a neuron in my brain whose job it is to ``know,'' say, Halle Berry \cite{Martindale}?

Strict connectionists argue against this interpretation \cite{CS}.  Instead, they claim, neurons that seem highly specific toward a stimulus only exhibit this specificity within the context of their networks.  Progressively higher-level features are extracted at each layer of processing as information about a stimulus percolates through the network.

If we take a neuron from an ANN and look under the hood, we see only the machinery described in equation \eqref{ANN}.  In this regard, all neurons in such a network are essentially alike and only give the suggestion of meaning by their activity within the complete network.  In short, the connectionist story goes, an \emph{isolated} neuron knows nothing.

\begin{figure}[t]
\begin{center}
\begin{tabular}{c}
\includegraphics[scale=0.5]{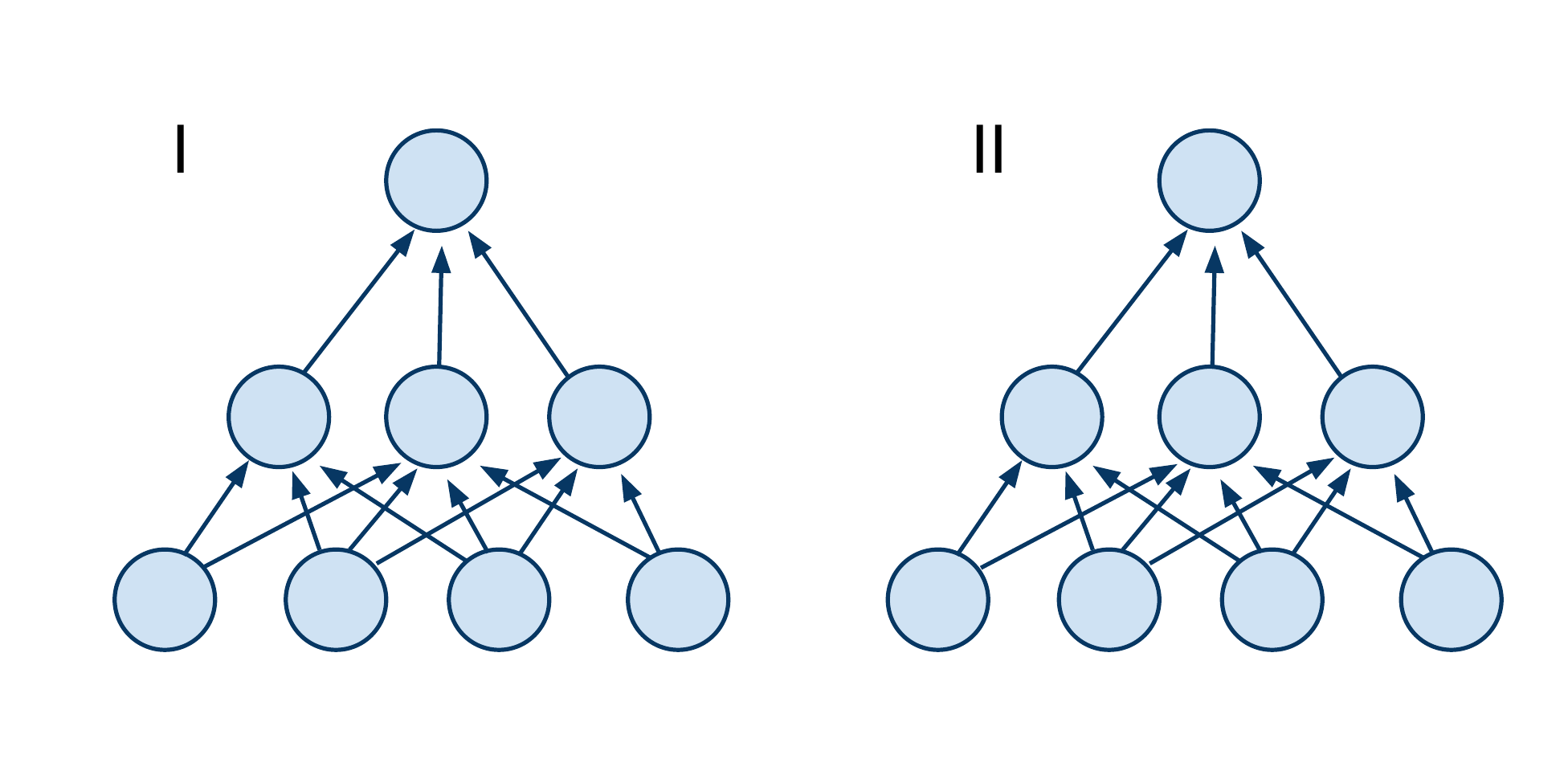} \\ \includegraphics[scale=0.5]{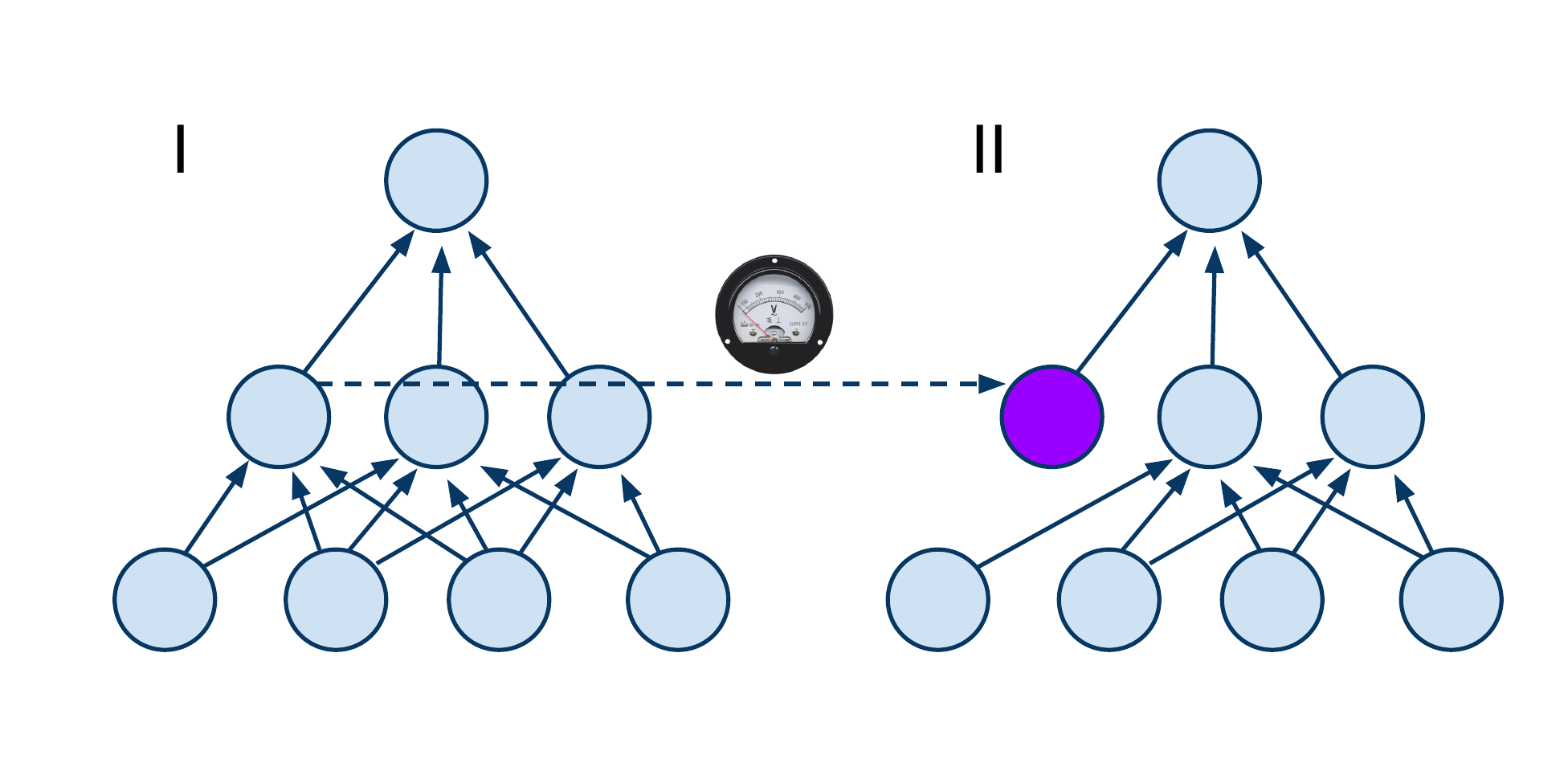}
\end{tabular}
\caption{Two identical networks (top) and the same networks following surgery and ``voltage linking'' (bottom).}
\end{center}
\end{figure}

If membrane potential is the variable of interest in cognitive information exchange, then it should be held to one of the standards that candidate neurotransmitters are: Namely, the cell should not really care where the voltage changes are coming from; it should only care that they are there.

Consider the two identical networks shown at the top of Figure 2.  One is a clone of the other, and we will imagine that every last molecular detail is the same in Network I ($N_{\text{I}}$) as in Network II ($N_{\text{II}}$).  Clearly, as these networks are identical, their responses to the same stimuli will be identical.  If we consider the single neuron at the top of the network to be the \emph{output}, then we can write 
\begin{equation}
	[ \mathcal I(N_{\text{I}}) = \mathcal I(N_{\text{II}}) ] \implies [ \mathcal O(N_{\text{I}}) = \mathcal O(N_{\text{II}}) ], \label{logic_1}
\end{equation}
which simply states symbolically what we have already said.  Namely, if the information, $\mathcal I,$ spread over the network is the same, then the output, $\mathcal O,$ will be the same also.\footnote{We cannot claim the converse, however, which says that the same output implies the same input.  Consider a network whose job it is to separate even and odd numbers.  The same output (e.g., ``even'') does not imply the same input (e.g., ``4'').}

Now let us perform surgery on the second network by removing the connections leading \emph{into} the leftmost neuron in the second layer, which we will call $\nu_5^{\text{II}}.$  We will make up for this insult, though, by connecting this neuron to its corresponding neuron, $\nu_5^{\text{I}},$ in the intact \emph{first} network and allow it to copy its membrane potential dynamics.  This arrangement is shown at the bottom of Figure 2.

If we now present the same stimuli to both networks, it is clear that the purple neuron in Network II should have identical firing behavior to its counterpart in Network I.  After all, we have set it up that way by ``voltage linking'' the two networks.  If the spike doctrine is correct, then all of the information in Network I is also present in Network II, since anything that was eliminated by the surgery is being imported into the network via our connecting device.  Even connection-strength changes, such as one would see in long-term potentiation or depression, only serve to affect the firing behavior and are therefore also accommodated for in this arrangement.

We have covered earlier (section 5.2) that a neuron's transmitter phenotype can change over time in response to environmental signals.  This process, however, is mediated by intracellular chemical signal transduction and gene expression, processes that are essentially independent of the membrane voltage.  If the signal we are feeding the networks produces a shift in transmitter phenotype in $\nu_5^{\text{I}},$ then even with an exchange of voltage information, $\nu_5^{\text{II}}$ will not show the same shift in transmitter phenotype.  Since the \emph{chemical} output from these corresponding cells no longer matches, there are at least some cases in which $\mathcal O(N_{\text{I}}) \neq \mathcal O(N_{\text{II}}),$ measured at the output neuron.

Returning to the relation in expression \eqref{logic_1}, we write its \emph{contrapositive}, which is true whenever expression \eqref{logic_1} is true,
\begin{equation}
	[ \mathcal O(N_{\text{I}}) \neq \mathcal O(N_{\text{II}}) ] \implies [ \mathcal I(N_{\text{I}}) \neq \mathcal I(N_{\text{II}}) ],
\end{equation}
and we must conclude that in the cases where the outputs do not match---which is a certainty in at least some cases involving a shift in transmitter phenotype---the information distributed over the network is not equivalent.  The only difference between the networks is chemical, so the specific \emph{chemistry} must account for the missing information.

This result holds irrespective of the method for evaluating the ``output'' of the network.  That is, one measure could be differences in firing patterns, while another could be the presence or absence of an IEG response within the cell.  In this case, however, since we are concerned with evaluating the spike doctrine, we will only look at the spiking behavior of the neuron.

Unless the outputs of the two networks are \emph{identical} in every way in every case, then it must be said that the voltage information cannot encode the stimulus but that the chemistry can.  And since we can be confident that changing the chemistry will change the firing behavior of downstream neurons, then we know the output will \emph{not} be identical.  Regardless of whether this network is supposed to respond to Halle Berry or the song of a zebra finch, it is clear that changing the chemistry changes the information distributed over the network.

Notice that earlier I said that the information would be different in \emph{some} cases instead of making the stronger claim that it would differ in \emph{all} cases.  That is because we must allow for the case in which a stimulus is such that \emph{stability} within the cell is maintained.  That is, some stimuli could conceivably be passed along without changing the cell's transmitter phenotype or gene-expression patterns.  In such stable cases, one \emph{would} obtain all the necessary information from observing the spike train, assuming that the \emph{current state} of the cell were known.

\section{Conclusion}

It is the position of this paper that a certain ``electrical chauvinism'' pervades modern cognitive science, an assumption that may have us looking in the wrong place in our search for \emph{information} in the brain.  If the main claim of this paper is correct, then it is the chemistry of the brain, and not the spiking of neurons, that defines the elementary units of information relevant to cognition.  It would hardly mean, however, that connectionism is dead.

Rather, the cognitive and computational sciences should look forward to an era of \emph{superconnectionism}, in which each node of a Turing-universal neural network contains \emph{another} Turing-universal network---a massively interacting chemical-genetic one---with vastly different properties and dynamics.

Although this does make the \emph{modeling} landscape considerably more complex, it redefines the problem in a way that allows us to reconsider the variables of computational importance in the cognitive system, variables whose very nature could have profound theoretical implications for the capabilities of the human brain.

Despite the appearance of additional complexity, there is an elegant \emph{simplicity} inherent in the brain's modern application of a chemical alphabet whose first words were spelled out in primordial pools billions of years ago.  To be sure, modeling this molecular language poses a significant challenge.  But if the language the brain speaks \emph{is} a chemical one---indeed, one having many dialects---then it is a challenge we must accept.


\begin{thebibliography}{99}

\bibitem{Bhalla2002}
Bhalla, U.S. (2002). ``Biochemical Signaling Networks Decode Temporal Patterns of Synaptic Input.''  \textit{Journal of Computational Neuroscience, 13}, 49--62.

\bibitem{Black}
Black, I.B. (1991). \textit{Information in the Brain: A Molecular Perspective}.  Cambridge, MA: The MIT Press.

\bibitem{Branco}
Branco, T., and Staras, K. (2009). ``The Probability of Neurotransmitter Release: Variability and Feedback Control at Single Synapses.''  \textit{Nature Reviews: Neuroscience, 10}, 373--383.

\bibitem{Bullock}
Bullock, T.H., \textit{et al.} (2005).  ``The Neuron Doctrine, Redux.''  \textit{Science, 310}, 791--793.

\bibitem{Butts}
Butts, D.A., \textit{et al.} (2007). ``Temporal Precision in the Neural Code and the Timescales of Natural Vision.''  \textit{Nature, 449}, 92--95.

\bibitem{Changeux}
Changeux, J.P. (1986).  ``Coexistence of Neuronal Messengers and Molecular Selection.'' \textit{Progress in Brain Research, 68}, 373--403.

\bibitem{CS}
Churchland, P.S., and Sejnowski, T.J. (1999).  \textit{The Computational Brain}.  Cambridge, MA: The MIT Press.

\bibitem{Clayton}
Clayton, D.F. (2000). ``The Genomic Action Potential.''  \textit{Neurobiology of Learning and Memory, 74}, 185--216.

\bibitem{Coolen}
Coolen, A.C.C., K\"{u}hn, R., and Sollich, P. (2005).  \textit{Theory of Neural Information Processing Systems}.  Oxford, UK: Oxford University Press.

\bibitem{DA}
Dayan, P., and Abbott, L.F. (2001).  \textit{Theoretical Neuroscience: Computational and Mathematical Modeling of Neural Systems}.  Cambridge, MA: The MIT Press.

\bibitem{Diamant} Diamant, E. (2008). ``Unveiling the mystery of visual information processing in human brain.'' \textit{Brain Research, 1225}, 171--178.

\bibitem{Dong}
Dong, S., \textit{et al.} (2009).  ``Discrete Molecular States in the Brain Accompany Changing Responses to a Vocal Signal.''  \textit{Proceedings of the National Academy of Sciences, 106(27)}, 11364--11369.




\bibitem{GIM}
Gazzaniga, M.S., Ivry, R.B., and Mangun, G.R. (2009).  \textit{Cognitive Neuroscience: The Biology of the Mind, Third Edition}.  New York, NY: Norton.

\bibitem{GK2002}
Gerstner, W., and Kistler, W.M. (2002). \textit{Spiking Neuron Models: Single Neurons, Populations, Plasticity}.  New York, NY: Cambridge University Press.

\bibitem{Gold}
Gold, I., and Stoljar, D. (1999).  ``A Neuron Doctrine in the Philosophy of Neuroscience.''  \textit{Behavioral and Brain Sciences, 22}, 809--869.


\bibitem{HK2001}
Ham, F.M., and Kostanic, I. (2001). \textit{Principles of Neurocomputing for Science and Engineering}.  New York, NY: McGraw-Hill.

\bibitem{Harvey}
Harvey, R.L. (1994). \textit{Neural Network Principles}.  Englewood Cliffs, NJ: Prentice-Hall.

\bibitem{HW}
Hu, Y., and Wolfram, S. ``Random 3D Nearest Neighbor Networks,'' 
 \url{http://demonstrations.wolfram.com/Random3DNearestNeighborNetworks/}.

\bibitem{Igaz}
Igaz, L.M., \textit{et al.} (2004).  ``Gene Expression During Memory Formation.''  \textit{Neurotoxicity Research, 6(3)}, 189--204.

\bibitem{Kandel}
Kandel, E.R., Schwartz, J.H., and Jessell, T.M. (2000).  \textit{Principles of Neural Science, Fourth Edition}.  New York, NY: McGraw-Hill.

\bibitem{KC}
Katz, P.S., and Clemens, S. (2001). ``Biochemical Networks in Nervous Systems: Expanding Neuronal Information Capacity Beyond Voltage Signals.''  \textit{TRENDS in Neurosciences, 24(1)}, 18--25.

\bibitem{Koch}
Koch, C. (1999).  \textit{Biophysics of Computation}.  New York, NY: Oxford University Press.

\bibitem{KE}
Koppell, N., and Ermentrout, B. (2004). ``Chemical and Electrical Synapses Perform Complementary Roles in the Synchronization of Interneuronal Networks.''  \textit{Proceedings of the National Academy of Sciences, 101(43)}, 15482--15487.

\bibitem{Lennie}
Lennie, P. (2003).  ``The Cost of Cortical Computation.''  \textit{Current Biology, 13}, 493--497.

\bibitem{Le Novere}
Le Nov\`{e}re, N. (2007).  ``The Long Journey to a Systems Biology of Neuronal Function.''  \textit{BMC Systems Biology, 1:28}.

\bibitem{Linden}
Linden, D.J. (2007).  \textit{The Accidental Mind: How Brain Evolution Has Given Us Love, Memory, Dreams, and God}.  Cambridge, MA: Harvard University Press.


\bibitem{Maass}
Maass, W. (1998). ``A Simple Model for Neural Computation with Firing Rates and Firing Correlations.'' \textit{Network: Computation in Neural Systems, 9}, 381--397.

\bibitem{Magnasco}
Magnasco, M.O. (1997).  ``Chemical Kinetics is Turing Universal.''  \textit{Physical Review Letters, 78(6)}, 1190--1193.

\bibitem{Martindale}
Martindale, D. (2005). ``One Face, One Neuron: Storing Halle Berry in a Single Brain Cell.''  \textit{Scientific American, 293(4)}. 

\bibitem{MP}
McCulloch, W.S., and Pitts, W. (1943).  ``A Logical Calculus of the Ideas Immanent in Nervous Activity.''  \textit{Bulletin of Mathematical Biophysics, 5}, 115--133.

\bibitem{Potter}
Potter, D.D., \textit{et al.} (1986).  ``Transmitter Status in Cultured Sympathetic Principal Neurons: Plasticity, Graded Expression and Diversity.'' \textit{Progress in Brain Research, 68}, 103--120.

\bibitem{Revest}
Revest, P. and Longstaff, A. (1998). \textit{Molecular Neuroscience}.  New York, NY: Springer.

\bibitem{Rieke}
Rieke, F., Warland, D., de Ruyter van Steveninck, R., and Bialek, W. (1999). \textit{Spikes: Exploring the Neural Code}.  Cambridge, MA: The MIT Press.


\bibitem{Schratt06}
Schratt, G. M., \textit{et al.} (2006).  ``A Brain-Specific microRNA Regulates Dendritic Spine Development.'' \textit{Nature, 439},
283--289. 

\bibitem{Schratt09}
Schratt, G. (2009).  ``microRNAs at the Synapse.''  \textit{Nature Reviews: Neuroscience, 10(12)}, 842--849.

\bibitem{Schultzberg}
Schultzberg, M., \textit{et al.} (1986).  ``Coexistence During Ontogeny and Transplantation.''  \textit{Progress in Brain Research, 68}, 129--145.

\bibitem{Scott}
Scott, A. (2002).  \textit{Neuroscience: A Mathematical Primer}.  New York, NY: Springer.

\bibitem{SS91}
Siegelmann, H.T., and Sontag, E.D. (1991). ``Turing Computability with Neural Nets.''  \textit{Applied Mathematics Letters, 4(6)}, 77--80.

\bibitem{Stahl}
Stahl, S.M. (2000).  \textit{Essential Psychopharmacology: Neuroscientific Basis and Practical Applications, Second Edition}. Cambridge, UK: Cambridge University Press.

\bibitem{Thagard}
Thagard, P. (2002). ``How Molecules Matter to Mental Computation.''  \textit{Philosophy of Science, 69}, 429--446.

\bibitem{TLA}
Thagard, P., Litt, A., and Aubie, B. (2007).  ``Your Brain on Drugs: Neurotransmitters, Receptors, and the Mind-Body Problem.'' Unpublished manuscript.


\bibitem{Trappenberg}
Trappenberg, T.P. (2002).  \textit{Fundamentals of Computational Neuroscience}.  New York, NY: Oxford University Press.

\bibitem{Trudeau}
Trudeau, L.E., and Guti\'{e}rrez, R. (2007).  ``On Cotransmission \& Neurotransmitter Phenotype Plasticity.''  \textit{Molecular Interventions, 7(3)}, 138--146.

\bibitem{vBD2006}
von Bohlen und Halbach, O., and Dermietzel, R. (2006). \textit{Neurotransmitters and Neuromodulators: Handbook of Receptors and Biological Effects, Second Edition}. Weinheim, Germany: Wiley-VCH Verlag GmbH.

\bibitem{Warren}
Warren, W.C., \textit{et al.} (2010). ``The Genome of a Songbird.''  \textit{Nature, 464}, 757--762.

\end{thebibliography}
\end{document}